# Properties and influence of microstructure and crystal defects in $Fe_2VAl$ modified by laser surface remelting

Leonie Gomell[1*], Moritz Roscher[1], Hanna Bishara[1], Eric Jägle[1,2], Christina Scheu[1], Baptiste Gault[1,3]*

[1] Max-Planck-Institut für Eisenforschung GmbH, Max-Planck-Str. 1, 40237, Düsseldorf, Germany

[2] Universität der Bundeswehr München, Institut für Werkstoffkunde, Werner-Heisenberg-Weg 39, 85577 Neubiberg

[3] Department of Materials, Royal School of Mines, Imperial College London, London, UK

;  * corresponding authors: l.gomell@mpie.de, b.gault@mpie.de

**Abstract**

Laser surface remelting can be used to manipulate the microstructure of cast material. Here, we present a detailed analysis of the microstructure of $Fe_2VAl$ following laser surface remelting. Within the melt pool, elongated grains grow nearly epitaxially from the heat-affected zone. These grains are separated by low-angle grain boundaries with 1°–5° misorientations. Segregation of vanadium, carbon, and nitrogen at grain boundaries and dislocations is observed using atom probe tomography. The local electrical resistivity was measured by an *in-situ* four-point-probe technique.  A smaller increase in electrical resistivity is observed at these low-angle grain boundaries compared to high-angle grain boundaries in a cast sample. This indicates that grain boundary engineering could potentially be used to manipulate thermoelectric properties.

Environmentally-friendly power generation is a major societal challenge currently faced by scientists. Research interest in thermoelectric (TE) materials is hence on the rise (again). Thermoelectricity is caused by the Seebeck effect [1]. While the physical background is well laid out, several strategies to increase efficiency have remained unexplored. The TE performance can be described by the figure of merit $zT$ ($zT = S^2\sigma/\kappa$), where $S$ is the Seebeck coefficient, $\sigma$ is the electrical conductivity, $S^2\sigma$ is the power factor, and $\kappa$ is the thermal conductivity. The high price or the toxicity of the elements constituting the materials have hindered widespread applications. Hence, a non-toxic and cost-effective alternative is needed.

A promising candidate is the full Heusler compound $Fe_2VAl$. The power factor of $Fe_2VAl$ is high (>5 mWm$^{-1}$K$^{-2}$ at 300 K) [2] and can compete with state-of-the-art materials such as $Bi_2Te_3$ [3,4]. Yet, the high thermal conductivity $\kappa$ restricts its thermoelectric performance. Therefore, the main aim is to increase phonon scattering to reduce $\kappa$. Possible approaches are doping [5–7], off-stoichiometric compositions [8–11], introduction of defects and disorder [12–14], and interfaces [15,16]. The microstructure can be manipulated by adjusting the synthesis route. A powerful way to change and control the microstructure is laser surface remelting (LSR) [17]. A focused laser beam is scanned over the surface, leading to a fast melting and subsequent solidification. The laser power and the scan velocity can be changed to adjust the peak temperature, the solidification speed, and thus, the resulting microstructure. LSR ensures a high quenching rate on the order of $10^3$-$10^8$ K/s [18], which depends on the scanning parameters. The parameters used here are expected to lead to an approximate quenching rate of $10^4$-$10^5$ K/s.

Here, we focus on the microstructure of $Fe_2VAl$ following LSR and its local influence on the electrical resistivity. Emphasis is placed on dislocations and low-angle grain boundaries (LAGB), and associated solute segregation. Local measurements of the electrical resistivity show that LAGBs do not significantly affect the electrical resistivity compared to high-angle GBs (HAGBs) in the $Fe_2VAl$ system. Our study advances the understanding of the microstructure – property relationship in these alloys, offering opportunities for further optimization of the thermoelectric performance.

Stoichiometric amounts of pure Fe (99.9%, Carboleg GmbH), Al (99.7%, Aluminium Norf), and V (99.9%, HMW Hauner GmbH) were arc-melted under an argon atmosphere. To ensure homogeneity, the sample was flipped over and remelted four times. The chemical composition obtained from inductively-coupled plasma optical emission spectrometry (ICP-OES) is $Fe_{50.09}V_{24.95}Al_{24.96}$ (at.%). 0.03 at.% C and 0.01 at.% N were measured by infrared absorption measurement and melting under helium and thermal conductivity measurement, respectively. Before LSR, this sample was hand-ground to 600 grit SiC paper. An AconityMini system by Aconity3D, equipped with an ytterbium-fiber laser with a wavelength of 1070 nm, was used to remelt the surface in argon atmosphere with an oxygen concentration below 80 ppm. The focal size of the laser beam was 90 µm, the laser power $p$ was 200 W, the scanning speed $v$ was 350 mm/s, and the hatch distance $h$ was 0.1 mm. The energy density is 6.67 J/mm². Microstructural analysis was performed using scanning electron microscopy (SEM) in backscattered electron mode (BSE). The cross-section was mechanically polished down to 0.05 µm colloidal silica. Imaging, electron backscattered diffraction (EBSD), and energy-dispersive X-ray spectroscopy (EDX) were conducted using a Zeiss 1540 XB SEM. EBSD was performed with a step size of 200 nm and an acceleration voltage of 15 kV. Needle-shaped specimens for atom probe tomography (APT) were prepared using a dual-beam focused-ion-beam (FIB) instrument (FEIHelios Nanolab600/600i) with a Ga ion source as described in Ref. [19]. APT specimens were analyzed using a LEAP™ 5000 XS (Cameca Instruments) operated in laser pulsing mode using a pulse repetition rate of 200 kHz and a pulse energy of 40 pJ. The base temperature of the specimen was kept at 60 K and the target detection rate was set to 4%. Data reconstruction and analysis were done in IVAS 3.8.4 and Matlab R2019.

The influence of LAGBs on electronic transport was studied by local electrical resistivity measurements carried-out *in-situ* inside an SEM (Zeiss Gemini) with a linear 4-point-probe technique. The sample was inserted into an SEM on a stage with four independent micromanipulators (PS4, Kleindiek Nanotechnik GmbH). Each micromanipulator contains a needle with a tip radius of approx. 50 nm. The needles are positioned 5 µm apart inside a single grain and then on either side of a GB. The contribution of an individual GB to the resistivity is the ratio between bi- and single-crystal regions. Technical details on the electrical measurements, e.g. errors, corrections due to deviation from equidistance positions of needles and modulation of electrical current, are detailed in Ref. [20].

Supplementary Fig. S1 gives an overview of the microstructure of the arc melted sample, with a grain size of (300 ± 45) µm. EDX indicates a homogenous composition on a micrometer level. EBSD reveals that grains are separated by HAGBs with no noticeable texture. Fig 1. shows optical and BSE images of the remelted area, in the top and the cross-sectional views, respectively. Cracks are observed (Fig. 1a/b) and mainly appear at the (high-angle) grain boundaries of the arc melted sample (Fig. S1a). A similar cracking behavior was observed in superalloys [21,22]. The melt pools are approximately 106 µm deep, 160 µm wide, and show an overlap of two adjacent melt pools of 30 µm. The high magnification BSE image shown in Fig. 1c reveals a structure with elongated solidification cells originating from within the melt pool, surrounded by a heat-affected zone. These cells, i.e. individual grains, grew (nearly) epitaxially and hence share nearly the same orientation, and are separated by LAGBs, as discussed below. This low misorientation between grains leads to a notable contrast in the BSE image. The elongated grains are approximately 1 µm x 10 µm, as measured in the center of the melt pool 5 µm away from the surface. Some grains span the entire melt pool from the heat-affected zone to the surface. The average grain size decreases towards the center and the top of the remelted track due to the increased solidification rate, as reported in the welding and laser surface processing literature [23–25]. No precipitates are observed at this scale.

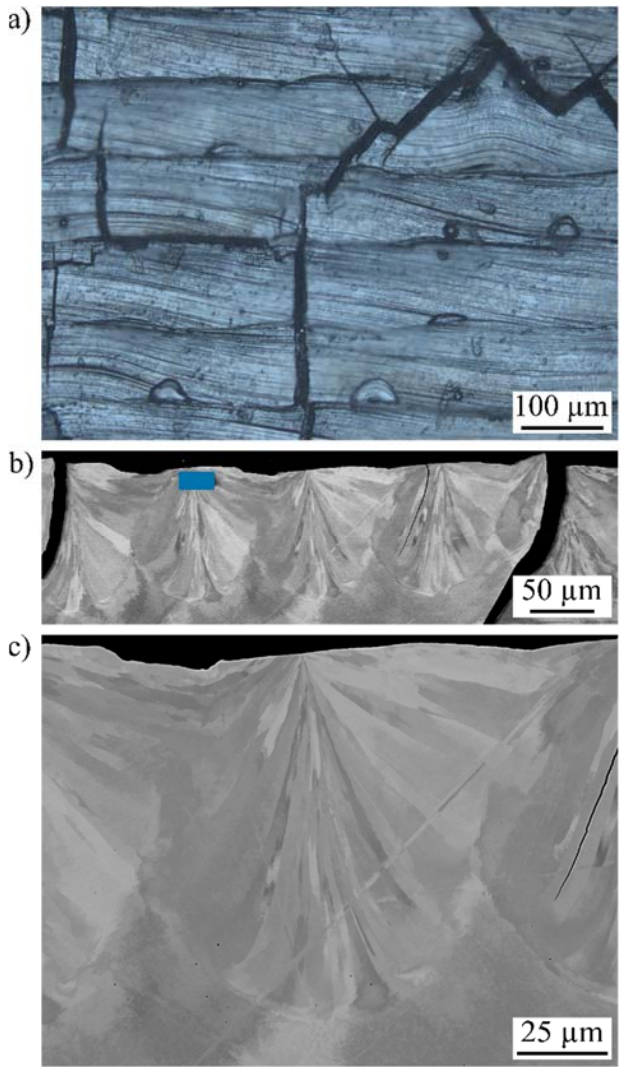

Fig 1a) Optical image of the surface of $Fe_2VAl$ treated with surface laser remelting. b) BSE image of the cross section of several melt pools with low magnification. The blue box marks the area of the APT lift out (not to scale). Cracks are visible, which appear mainly along grain boundaries of the arc melted sample. The remelting process was conducted from the right to the left side of the image. c) BSE image of the melt pool with a higher magnification. Elongated grains, growing from the HAZ to the impact point of the laser are observed.

EDX maps show a homogeneous distribution across the melt pool and surroundings (Supplementary Fig. S2). No preferential evaporation of an element occurred during remelting. EBSD was performed to investigate the crystallographic orientation differences within the melt pool (Fig. 2). At room temperature, $Fe_2VAl$ crystallizes in the $L2_1$ full Heusler phase [26]. With EBSD, all points could be indexed to this phase. The inverse pole figure (Fig. 2a) shows little orientation differences within the grain structure. The [001] reference axis is chosen with regard to the direction of the laser beam (x, A1), the direction of the arc melted surface (y, A2), and the scanning direction of the laser beam (z). No HAGBs (misorientation higher than 10°) are observed. The substrate grain is approximately along the [122] orientation. The different grains are oriented between [112] and [122] orientation, confirming epitaxial or near-epitaxial growth between the melt pool and substrate, as previously observed in LSR [27], selective laser melting [21], or laser powder-bed fusion [28,29]. In larger melt pools, HAGBs with a favored growth direction are observed within the melt pool [29], which could not be found in the present study. The misorientation between the second nearest neighbor is displayed in the Kernel map (Fig 2b). Such LAGBs are characterized by a high density of geometrically necessary dislocations (GNDs). The fraction of grain boundaries with different misorientations was calculated and is plotted in Fig. S3 (Suppl. Information). Nearly 50% of all observed boundaries show misorientations of 1° ± 0.25°. The fraction of boundaries with higher misorientations decreases towards a misorientation angle of 10°. These higher misorientations, around 10°, are observed close to the laser incidence position at the surface.

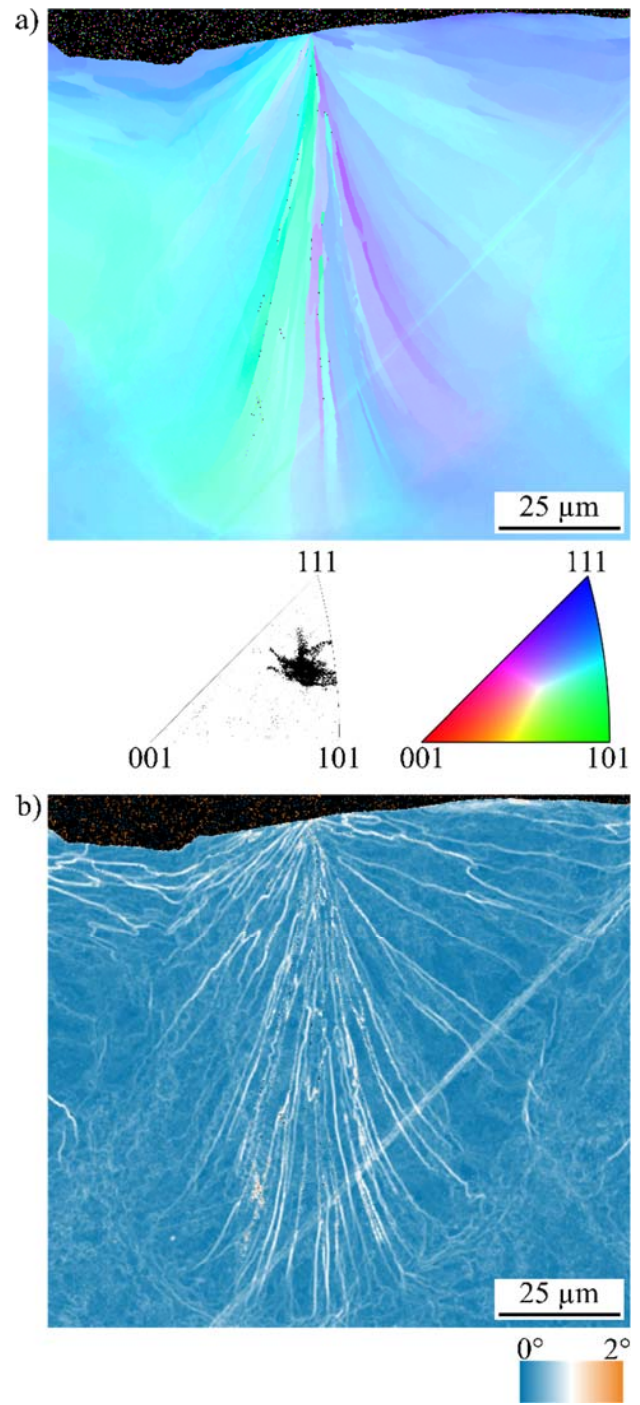

Fig 2a) EBSD inverse pole figure of the melt pool shown in Fig. 1c. Low orientation differences are found. The [001] reference axis is chosen in the laser scanning direction (z). The sample is orientated such that the remelted surface is parallel to y(A2), and the laser beam is parallel to x(A1). b) EBSD kernel map showing the misorientations of the 2$^{nd}$ nearest neighbor. The scale is set for misorientations between 0° and 2°.

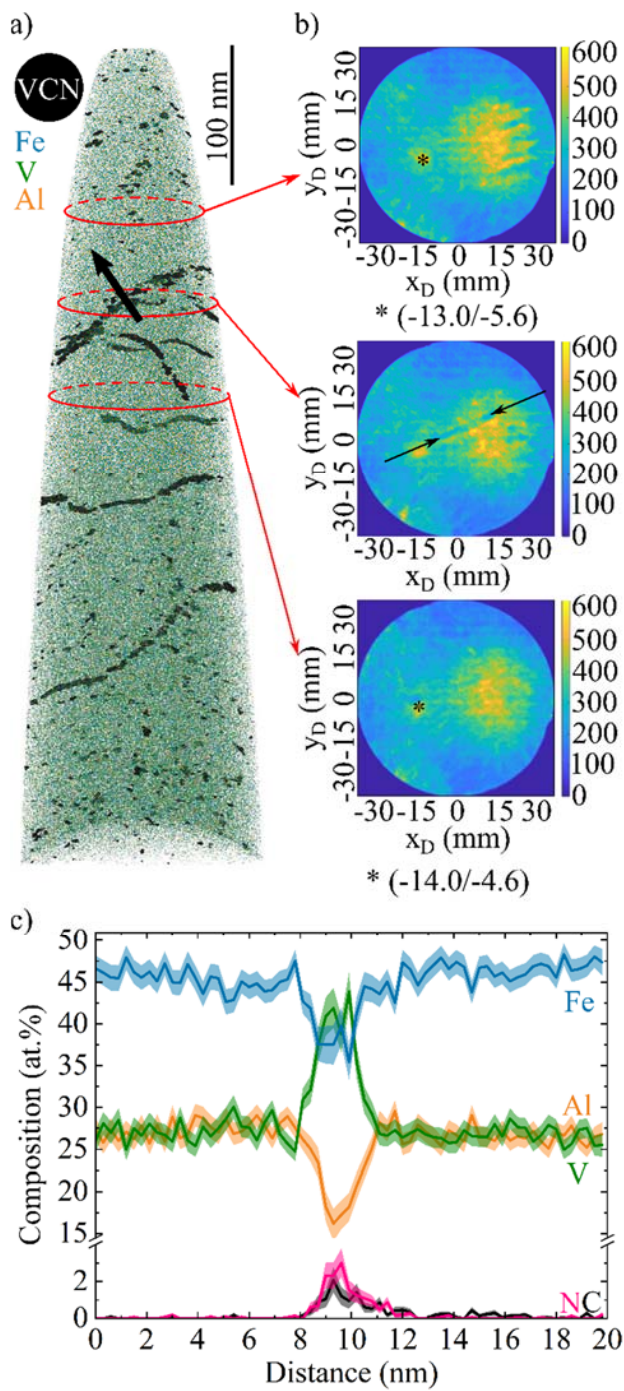

Fig 3a) APT reconstruction showing Fe, V, Al ions as blue, green, and orange dots, respectively. A black iso-composition surface for 0.5 at.% (VC, VN, C, N) is shown to visualize precipitation at a LAGB and dislocations. b) Detector hit maps at the boundary and on either side of the boundary. The boundary is visible in the second map, indicated by the black arrows. The position of the (111) pole on either side of the boundary is marked by an asterisk and given below the maps in detector coordinates. The position can be used to determine the misorientation angle of the boundary to be approximately 1.1°. c) composition profile along the low angle boundary, marked by the black arrow in a). Please note the different scales on the y-axis for the different regions. An enhancement of V, N, and C is found at the grain boundary, indicating $VC_xN_y$ precipitates.

We used APT in search for possible segregation or clustering within the melt pool that can affect TE properties. Specimens were prepared from a region close to the surface of the melt pool, as marked by the blue box (not to scale) in Fig. 1b. Fig. 3a shows a representative APT reconstruction. The Fe, V, and Al ions are displayed in blue, green, and orange, respectively. The black iso-composition surface delineates regions with a composition above 0.5 at.% VN, VC, N, and C ions combined. A boundary and several dislocations are observed, covered with $VC_xN_y$ precipitates. The dislocations are observed near the boundary and in the bulk. The dislocation density is moderate with an approximate distance of less than 100 nm between dislocations.

Three APT detector hit maps on either side and at the boundary were plotted (Fig. 3b), each showing the (111) pole. The (011) pole is also visible but more difficult to locate precisely. Two more {011} poles are slightly outside the detector map. The second map shows a darker feature, indicated by the black arrows, which corresponds to the grain boundary. The distance between the poles on either side of the boundary is 1.4 mm, which corresponds to a misorientation of the LAGB of approximately 1.1° assuming that the field-of-view is close to 60° on this instrument [30]. This LAGB has a misorientation similar to those determined by EBSD.

A composition profile was calculated along a cuboidal region-of-interest (5x13 nm$^2$) with a step size of 0.4 nm (Fig. 3c) across the LAGB. The position of the profile is indicated by the black arrow in Fig. 3a. This compositional plot is exemplary for the compositional changes across the boundary and the different dislocations. More composition profiles are given in the supplementals (Fig. S4). At the boundary and the dislocations, the composition is $Fe_{37.5\pm1.5}V_{41.4\pm1.8}Al_{17.5\pm0.8}C_{1.3\pm0.4}N_{2.3\pm0.5}$ (at.%). Averaged over a 500 million ion data set, with the error corresponding to the 2σ of the counting statistic, the composition is $Fe_{49.71\pm0.01}V_{26.26.4\pm0.1}Al_{24.00\pm0.01}C_{0.013\pm0.001}N_{0.016\pm0.005}$ (at.%). This value is close to the bulk composition of the arc melted sample measured by ICP-OES. No indication for grain to grain variations could be observed, including across different APT specimens.

The observed precipitates are comparable to those imaged by SEM in melt spun $Fe_2VAl$ [15] that were 200 to 300 nm wide and few nanometers thick. After LSR no precipitation is observed by SEM, only revealed by APT due to their small sizes. The composition of C and N is lower than in this previous report. This difference can be attributed to a different purity of the synthesis processes. Vanadium is known for its high affinity to form carbides and nitrides. C and N might be more available in the atmosphere of the melt spinning process compared to arc melting and LSR. While melt spinning results in spherical grains with HAGBs, LSR results in elongated grains with LAGBs. The different boundaries can influence precipitation and the level of segregation due to a different driving force, i.e. boundary energy.

The resistivity was analyzed within the melt pool at two different LAGBs distinguished by misorientation, and within a bulk HAGB unaffected by the laser remelting. The measurement positions are indicated in the supplementary Fig. S5a/b. To obtain absolute values of resistivity from the measured sheet resistance, correction factors are required [31]. However, the irregular shape of the remelted surface prevents deriving corresponding correction factors and consequently resolving the resistivity within the melt pool, to inspect if it is affected by the relatively high GNDs density. Nevertheless, the relative increase in resistivity $\tilde{R}$ due to a single GB is obtained as the ratio between resistance across GB and resistance within a single grain. This ratio is reliable when the positions of needles for measuring single- and bi-crystals are close enough to each other (less than the distance between the probing needles). Then the correction factor is similar for both of them, thereby canceling out in the ratio. A possible resistivity variation between the bulk and the melt pool is eliminated by using the ratio. Therefore, $\tilde{R}$ yields the direct contribution of the GB to resistivity of the material. For each boundary, $\tilde{R}$ and the corresponding misorientations determined by EBSD are given in Tab. 1. We consider the rotation angle as the main GB structural parameter due to the uniform out-of-plane directions within the melt pool. At the HAGB, the resistivity is enhanced by a factor of 1.14 compared to the neighboring grains due to electron scattering. In contrast, the LAGBs show a resistivity enhancement of 1.04-1.07. These local electrical measurements reveal a reduced influence of the LAGBs within the melt pool on resistivity compared to the HAGBs within the original sample. Despite the known dependence of electron scattering with GB structure [20,32], our results are a direct evidence of the minor effect of GBs within the melt pool on the resistivity, relatively to the original materials' GBs. Yet, the grain size within the melt pool is significantly reduced and thus the relatively high density of LAGB will contribute to an overall enhancement of the resistivity.

Tab. 1: Misorientation and relative increase in the resistivity $\tilde{R}$ due to a single GB. The positions of the boundaries are shown in the supplementary Fig. S5.

|  | GB1 | GB 2 | GB 3 |
|---|---|---|---|
| Mis-orientation | 2.5° ± 0.3° | 5.3° ± 0.3° | 20.8° ± 0.5° |
| $\tilde{R}$ | 1.04 ± 0.03 | 1.07 ± 0.03 | 1.14 ± 0.03 |

Enhancing thermoelectric performance requires minimizing the product of the electrical resistivity and the thermal conductivity. Thus, the enhanced electrical resistivity can be compensated by increased phonon scattering at the boundaries. At room temperature, the mean free path of the electrons is typically on the order of several nanometers, while for phonons it is approximately 2.4 μm. The impact of grain boundaries of a high-density of 1 μm-grains is assumed to have a stronger influence on the phonons than on the electrons. However, the influence of the observed microstructure on the thermal conductivity could not be analyzed here. In the literature, the impact of dislocations and LAGBs has been investigated on $Bi_2Te_3$ [33,34] and half-Heusler compound NbCoSn [35]. In these studies, the TE figure of merit could be enhanced by grain refinement and a high dislocation density [35]. The impact of the phonon scattering was large and could outperform the degraded electrical conductivity.

Due to the low electron scattering investigated here at LAGBs and the hypothesized phonon scattering, we can assume that LSR enhances the thermoelectric properties compared to arc melting.

In conclusion, we investigated the microstructure and local electrical resistivity of $Fe_2VAl$ treated with laser surface remelting. Within the melt pool, we observed a high density of dislocations and small elongated grains, separated by LAGBs with typical misorientation of 1°–5°. Parallel to the surface, the grain size is approximately 1 µm and therefore smaller than the mean free path of the phonons (approx. 2.4 µm). APT also revealed segregation of vanadium, carbon, and nitrogen towards these defects, likely stabilizing them by lowering their free energy. The electrical resistivity change at a LAGB is lower than at a HAGB. Together with the assumed enhancement of the phonon scattering due to the LAGBs, the thermoelectric properties are expected to be enhanced by changing the synthesis route to LSR. The flexibility of LSR, i.e. in terms of an adjustable cooling rate, enables further investigations of the impact of the cooling rate on defects and thermoelectric properties.

**Acknowledgement**

L. G. gratefully acknowledges IMPRS-SurMat and Studienstiftung des deutschen Volkes for funding. U. Tezins, A. Sturm, C. Broß, M. Nellessen, and K. Angenendt are acknowledged for their technical support at the FIB/APT and SEM facilities at MPIE. H.B. acknowledges the financial support by the ERC Advanced Grant GB CORRELATE (Grant Agreement 787446 GB-CORRELATE) of Gerhard Dehm.

**References**

[1]   T.J. Seebeck, in: J.C. Poggendorff (Ed.), Ann. Phys., Verlag von Joh. Ambrosius Barth, Leipzip, 1826.
[2]   E. Alleno, Metals (Basel). 8 (2018) 864.
[3]   S.K. Mishra, S. Satpathy, O. Jepsen, J. Phys. Condens. Matter 9 (1997) 461–470.
[4]   O. Yamashita, S. Tomiyoshi, K. Makita, J. Appl. Phys. 93 (2003) 368–374.
[5]   M. Mikami, Y. Kinemuchi, K. Ozaki, Y. Terazawa, T. Takeuchi, J. Appl. Phys. 111 (2012).

[6]     K. Renard, A. Mori, Y. Yamada, S. Tanaka, H. Miyazaki, Y. Nishino, J. Appl. Phys. 115 (2014).

[7]     Y. Terazawa, M. Mikami, T. Itoh, T. Takeuchi, J. Electron. Mater. 41 (2012) 1348–1353.

[8]     T. Takeuchi, Y. Terazawa, Y. Furuta, A. Yamamoto, M. Mikami, J. Electron. Mater. 42 (2013) 2084–2090.

[9]     M. Mikami, M. Inukai, H. Miyazaki, Y. Nishino, J. Electron. Mater. 45 (2016) 1284–1289.

[10]    Y. Nishino, S. Kamizono, H. Miyazaki, K. Kimura, AIP Adv. 9 (2019).

[11]    H. Miyazaki, S. Tanaka, N. Ide, K. Soda, Y. Nishino, Mater. Res. Express 1 (2014).

[12]    C. van der Rest, A. Schmitz, P.J. Jacques, Acta Mater. 142 (2018) 193–200.

[13]    S. Bandaru, A. Katre, J. Carrete, N. Mingo, P. Jund, Nanoscale Microscale Thermophys. Eng. 21 (2017) 237–246.

[14]    D.I. Bilc, P. Ghosez, Phys. Rev. B 83 (2011) 1–6.

[15]    L. Gomell, S. Katnagallu, A.D. Rasselio, S. Maier, L. Perrière, C. Scheu, E. Alleno, B. Gault, Scr. Mater. 186 (2020) 370–374.

[16]    S. Maier, S. Denis, S. Adam, J.C. Crivello, J.M. Joubert, E. Alleno, Acta Mater. 121 (2016) 126–136.

[17]    J. Zhang, B. Song, Q. Wei, D. Bourell, Y. Shi, J. Mater. Sci. Technol. 35 (2019) 270–284.

[18]    Y. Zhang, X. Lin, L. Wang, L. Wei, F. Liu, W. Huang, Intermetallics 66 (2015) 22–30.

[19]    K. Thompson, D. Lawrence, D.J. Larson, J.D. Olson, T.F. Kelly, B. Gorman, Ultramicroscopy 107 (2007) 131–139.

[20]    H. Bishara, M. Ghidelli, G. Dehm, ACS Appl. Electron. Mater. 2 (2020) 2049–2056.

[21]    P. Kontis, E. Chauvet, Z. Peng, J. He, A.K. da Silva, D. Raabe, C. Tassin, J.-J. Blandin, S. Abed, R. Dendievel, B. Gault, G. Martin, Acta Mater. 177 (2019) 209–221.

[22]    E. Chauvet, P. Kontis, E.A. Jägle, B. Gault, D. Raabe, C. Tassin, J.J. Blandin, R. Dendievel, B. Vayre, S. Abed, G. Martin, Acta Mater. 142 (2018) 82–94.

[23]    W. Kurz, R. Trivedi, Mater. Sci. Eng. A 179–180 (1994) 46–51.

[24]    V. Fallah, M. Amoorezaei, N. Provatas, S.F. Corbin, A. Khajepour, Acta Mater. 60 (2012) 1633–1646.

[25]    A. Frenk, W. Kurz, Mater. Sci. Eng. A 173 (1993) 339–342.

[26]    A. Berche, P. Jund, Comput. Mater. Sci. 149 (2018) 28–36.

[27]    M.N. Patel, D. Qiu, G. Wang, M.A. Gibson, A. Prasad, D.H. Stjohn, M.A. Easton, Scr. Mater. 178 (2020) 447–451.

[28]    M.-S. Pham, B. Dovgyy, P.A. Hooper, C.M. Gourlay, A. Piglione, Nat. Commun. 11 (2020) 749.

[29]    V. Ocelík, I. Furár, J.T.M. De Hosson, Acta Mater. 58 (2010) 6763–6772.

[30]  F. De Geuser, B. Gault, Microsc. Microanal. 23 (2017) 238–246.

[31]  I. Miccoli, F. Edler, H. Pfnür, C. Tegenkamp, J. Phys. Condens. Matter 27 (2015).

[32]  M. César, D. Liu, D. Gall, H. Guo, Phys. Rev. Appl. 2 (2014) 1–11.

[33]  S. Il Kim, K.H. Lee, H.A. Mun, H.S. Kim, S.W. Hwang, J.W. Roh, D.J. Yang, W.H. Shin, X.S. Li, Y.H. Lee, G.J. Snyder, S.W. Kim, Science (80-. ). 348 (2015) 109–114.

[34]  L. Yang, Z.G. Chen, M. Hong, G. Han, J. Zou, ACS Appl. Mater. Interfaces 7 (2015) 23694–23699.

[35]  G. Rogl, S. Ghosh, L. Wang, J. Bursik, A. Grytsiv, M. Kerber, E. Bauer, R.C. Mallik, X.Q. Chen, M. Zehetbauer, P. Rogl, Acta Mater. 183 (2020) 285–300.